\documentstyle[12pt]{article}
\newcounter{eqns}

\begin{document}
\title{Comment on Recent Argument That Neutrinos Are Not Majorana Particles%
    \thanks{National Science Foundation preprint NSF-PT-97-1. 
    The statements in this paper are not official views of the National Science Foundation.}}
\author{Boris Kayser\\ \\National Science Foundation\\
    4201 Wilson Boulevard\\Arlington, VA  22230  USA}
\date{February 1997}
\maketitle
\bigskip

\begin{quotation}
\begin{center} {\Large {\bf Abstract}} \end{center}
Existing data on neutrino-electron scattering do not imply that
neutrinos are not Majorana particles. The question of whether
neutrinos are of Majorana or of Dirac character remains completely open.
\end{quotation}

\noindent 
\rule{\textwidth}{.02pt}
\bigskip

Recently, it has been argued$^{\ref{ref1}}$ that existing data imply 
that neutrinos are not Majorana particles. The argument is as follows:
\begin{enumerate}
\item If some neutrino is a Majorana particle, then its vector neutral 
current (NC) vanishes.
\item The CHARM~II experiment has found that, at least at the $2\sigma$ 
level, the product of the vector NC couplings of a $\nu_\mu$ and an 
electron is nonzero.$^{\ref{ref2}}$ 
\item If, as suggested by the CHARM~II experiment, the $\nu_\mu$ 
vector NC coupling is really nonzero, then the $\nu_\mu$ cannot be a 
Majorana neutrino. In that case, the other neutrinos are probably 
not Majorana particles either.
\end{enumerate}

This argument is not correct. In particular, the assertion in step~(2) 
is not right. The CHARM~II experiment studied the NC reactions 
$\stackrel{\mbox{\tiny(---)\normalsize}}{\nu_\mu}\!e\! \rightarrow 
\stackrel{\mbox{{\tiny(---)}}}{\nu_\mu}\!\!e$. 
As explained below, experiments on NC interactions between a $\nu_\mu$ 
(and $\bar\nu_\mu$) and some target cannot determine the $\nu_\mu$ NC 
vector coupling.

In lowest order, the NC reactions $\stackrel{\mbox{{\tiny(---)}}}{\nu}\!+A 
\rightarrow \stackrel{\mbox{{\tiny(---)}}}{\nu}\!+B$ are, of course, due to 
Z boson exchange. Suppose the neutrino $\nu$ couples to the Z boson 
through the general V and A neutral current
\begin{equation}
J_\alpha^\nu = \bar\nu \gamma_\alpha (g_V^\nu + g_A^\nu \gamma_5) \nu.
\end{equation}
Here, the neutrino field $\nu$ may be either a Dirac field or a Majorana 
one. If it is a Majorana field, then, as Ref.~\ref{ref1} correctly notes, 
the vector neutral current $\bar\nu \gamma_\alpha \nu$ vanishes.

The amplitudes for $\nu+A \rightarrow \nu+B$ and $\bar\nu+A \rightarrow 
\bar\nu+B$ involve the neutrino neutral current $J_\alpha^\nu$ only through 
the matrix elements $\langle\nu_f|J_\alpha^\nu|\nu_i\rangle$ and 
$\langle\bar\nu_f|J_\alpha^\nu|\bar\nu_i\rangle$, respectively. Here, the 
subscripts $i$ and $f$ refer to the initial and final neutrino momenta 
and helicities. Now, in practice, the neutrinos and antineutrinos in our 
experimental beams are highly relativistic. In addition, the particles 
we call neutrinos are left-handedly polarized, while the ones we call 
antineutrinos are right-handedly polarized. These polarizations are, 
of course, observed facts, and do not depend on whether neutrinos are 
Dirac or Majorana particles. For example, the muons from $\pi^+ \rightarrow
\mu^+\nu_\mu$ are observed to be left-handed, from which it follows that 
the neutrinos are as well. For neutrinos and antineutrinos with the observed
polarizations, the matrix elements of the current $J_\alpha^\nu$ of Eq.~(1) 
are readily found to be:
\setcounter{eqns}{\theequation}
\addtocounter{eqns}{1}
\renewcommand{\theequation}{\arabic{eqns}.\arabic{equation}}
\usecounter{equation}
\begin{eqnarray}
\langle\nu_f|J_\alpha^\nu|\nu_i\rangle &=& c^\nu \overline{u_{fL}}
    \gamma_\alpha u_{iL} ,\\ [\jot]
\langle\overline{\nu_f}|J_\alpha^\nu|\overline{\nu_i}\rangle &=& -c^\nu 
\overline{v_{iR}}\gamma_\alpha v_{fR} . 
\end{eqnarray}
\setcounter{equation}{\theeqns}
\renewcommand{\theequation}{\arabic{equation}}
Here, $u$ and $v$ are the usual Dirac spinors, and the subscript $L(R)$ 
tells us that the particle described is left-handed (right-handed). 
In the Majorana case, ``$\overline{\nu_i}$'' and ``$\overline{\nu_f}$'' in 
Eq.~(2.2) are just the right-handed neutral particles produced, for example, 
in $\pi^-$ decay. Finally, the constant $c^\nu$ is given by
\begin{equation}
c^\nu = \left\{ \begin{array}{ccl}        
        g_V^\nu + g_A^\nu &;   & \nu\  \mbox{is a Dirac neutrino} \\ [\jot]
        2\,g_A^\nu             &;   & \nu\  \mbox{is a Majorana neutrino}
        \end{array}  \right.
\end{equation}
(We have omitted from $c^\nu$ an irrelevant overall constant.)

From Eqs.~(2), we see that the only thing the reactions 
$\stackrel{\mbox{{\tiny(--)}}}{\nu}\!+A 
\rightarrow \stackrel{\mbox{{\tiny(--)}}}{\nu}\!+B$ can teach us about 
the neutral current of $\nu$ is the value of the constant $c^\nu$. 
For a given value of this constant, the NC matrix elements of Eqs.~(2) 
are identical in the Majorana and Dirac cases. (The spinors $v$ in 
Eq.~(2.2) don't know whether the particles they are being used to 
describe are really ``antiparticles'' or not. These spinors are exactly 
the same quantities in either case.) Thus, the reactions 
$\stackrel{\mbox{{\tiny(--)}}}{\nu}\!+A 
\rightarrow \stackrel{\mbox{{\tiny(--)}}}{\nu}\!+B$ cannot tell us
whether $\nu$ is a Majorana particle or a Dirac one.

From Eq.~(3), we see that the relation between $c^\nu$ and the underlying 
NC coupling constants $g_V^\nu$ and $g_A^\nu$ does depend on whether $\nu$ 
is a Dirac neutrino or a Majorana one. However, since experiments on 
$\stackrel{\mbox{{\tiny(--)}}}{\nu}\!+A \rightarrow 
\stackrel{\mbox{{\tiny(--)}}}{\nu}\!+B$ measure only $c^\nu$, they 
cannot determine $g_V^\nu$ or $g_A^\nu$ separately. In particular, they 
cannot tell us that $g_V^\nu \neq 0$. It should also be noted that a 
theoretical model {\it can} contain a nonvanishing $g_V^\nu$ even though 
$\nu$ is a Majorana neutrino. This is true because in the Majorana case, 
the vector current $\bar\nu \gamma_\alpha \nu$ which multiplies $g_V^\nu$ 
in the neutral current $J_\alpha^\nu$ vanishes, so that $g_V^\nu$ does
not contribute to any measurable quantity.

Indeed, in the Standard Model (SM), a neutrino $\nu$ couples to the Z 
through a left-handed current. In the Sakurai gamma-matrix conventions 
which we are using here throughout, this means that in the SM, 
$g_V^\nu = g_A^\nu = 1/2$. This is true whether the mass terms added to
the ``minimal'' SM to give neutrinos masses make $\nu$ a Dirac particle 
or a Majorana one.

One might have wondered whether the measured value of $c^\nu$ can
tell us whether $\nu$ is of Dirac or Majorana character if we
are willing to assume the SM. Clearly, even if we do assume the
SM, the value of $c^\nu$ cannot distinguish between the Majorana
and Dirac cases. Indeed, from Eq.~(3), we see that the SM
predicts that $c^\nu=1$, regardless of whether $\nu$ is a Dirac
neutrino or a Majorana one.$^{\ref{ref3}}$

The argument of Ref.~\ref{ref1} seems to have grown out of one of those
misunderstandings which occur from time to time. The CHARM~II paper to which
Ref.~\ref{ref1} refers can be read as saying that (disregarding the errors) the
experiment has found that $g_V^{\nu_\mu}g_V^e \neq\ 0 $. However, this paper
refers in turn to an earlier one$^{\ref{ref4}}$ which rather clearly talks about
determining only an NC neutrino coupling $g^\nu$ with no $V$ or $A$ subscript.
That is absolutely right; as we have seen, NC experiments do not determine the $V,
A$ content of the neutrino neutral current.

\section*{References}
\begin{enumerate}
\item \label{ref1} R. Plaga, Max-Planck-Institut preprint, hep-ph/9610545.
\item \label{ref2} Here, the author of Ref.~\ref{ref1} refers to P. Vilain 
    et al.\ (CHARM~II Collaboration), {\it Phys.\ Lett.\/} {\bf B335}, 246 (1994).
\item \label{ref3} For further discussion of neutrino NC interactions and 
    the Majorana-Dirac distinction, please see B. Kayser and R. Shrock, 
    {\it Phys.\ Lett.\/} {\bf 112B}, 137 (1982), and B. Kayser, F. Gibrat-Debu and 
    F. Perrier, {\it The Physics of Massive Neutrinos }(World Scientific, 
    Singapore, 1989), Chap. 4.
\item \label{ref4} P. Vilain et al.\ (CHARM~II Collaboration), {\it Phys.\ Lett.\/} 
    {\bf B320}, 203 (1994).
\end{enumerate}

\end{document}